\documentclass[12pt,cites,graphicx]{article}
\usepackage{epsfig}

\title{Photoprocesses in protoplanetary disks}

\author{Ewine F.\ van Dishoeck,$^a$ Bastiaan Jonkheid$^a$ \\
 and Marc C.\ van Hemert$^{b}$\\[3mm]
$^a$ Leiden Observatory, P.O. Box 9513, 2300 RA Leiden,
The Netherlands. \\
E-mail: ewine@strw.leidenuniv.nl\\[3mm]
$^b$ Leiden Institute of Chemistry, P.O. Box 9502, \\
2300 RA Leiden, The Netherlands}

\date{\today}

\begin{document}
\maketitle
\renewcommand{\thefootnote}{\fnsymbol{footnote}}

%\begin{abstract}
\noindent Circumstellar disks are exposed to intense ultraviolet (UV)
radiation from the young star.  In the inner disks, the UV radiation
can be enhanced by more than seven orders of magnitude compared with
the average interstellar radiation field, resulting in a physical and
chemical structure that resembles that of a dense photon-dominated
region (PDR).  This intense UV field affects the chemistry, the
vertical structure of the disk, and the gas temperature, especially in
the surface layers. The parameters which make disks different from
more traditional PDRs are discussed, including the shape of the UV
radiation field, grain growth, the absence of PAHs, the gas/dust ratio
and the presence of inner holes. Illustrative infrared spectra from
the Spitzer Space Telescope are shown. New photodissociation cross
sections for selected species, including simple ions, are
presented. Also, a summary of cross sections at the Lyman $\alpha$
1216 \AA \ line, known to be strong for some T Tauri stars, is
made. Photodissociation and ionization rates are computed for
different radiation fields with color temperatures ranging from 30000
to 4000 K and grain sizes up to a few $\mu$m. The importance of a
proper treatment of the photoprocesses is illustrated for the
transitional disk toward HD 141569A which includes grain growth.
%\end{abstract}

\section{Introduction}
\label{intro}
It is well established observationally and theoretically that many
young stars in the solar neighborhood are surrounded by disks, i.e.,
flattened rotating concentrations of gas and dust (see \cite{Greaves05}
for a review). In the early phases, up to a few Myr after collapse of
the parent cloud, the disks are rich in gas and dust inherited from
the cloud core. Once accretion onto the star stops and the disk
becomes less turbulent, the $\sim$0.1 $\mu$m interstellar grains
coagulate to larger and larger particles and settle to the midplane,
eventually forming km-sized planetesimals which interact
gravitationally to form protoplanets (e.g., \cite{Weiden97}). In
this period, between a few and 10 Myr, also most of the gas may be
dissipated from the disk although the precise time scale for gas
removal is still uncertain. Dust disks are observed around much older
(up to a few Gyr) stars as well, but the grains in these disks are the
`debris' produced by collisions and fragmentation of planetesimals and
are no longer the original interstellar particles (e.g., \cite{Lagrange00}).
This paper is concerned with young, gas-rich disks up to
the transitional stage when the disk becomes optically thin to UV
radiation.

Young stars are known to be powerful emitters of UV radiation, with
typical intensities at the disk surface that are orders of magnitude
higher than the average interstellar radiation field \cite{Herbig86}.
For some sources, most of the flux may be contained in the Lyman
$\alpha$ line \cite{Bergin03}. These UV photons play a very important
role in the physical and chemical structure of the disk, especially in
the surface and intermediate layers.  Many disks are thought to have a
flaring structure, in which the disk height $H$ and opening angle
$H/R$ increase with radius $R$ (e.g., \cite{Chiang97,Alessio98}). In
such a geometry the disk surface intercepts significantly more UV
radiation from the star than in a flat geometry, heating the surface
layers and increasing the scale height further when hydrostatic
equilibrium is assumed in the vertical direction.  Because the
density in the upper layers is below that at which gas-grain coupling
becomes effective, the gas temperature can be higher than the dust
temperature due to photoelectric heating, affecting both the
structure, chemistry and line formation \cite{Jonkheid04,Kamp04a}.  In
the inner disk ($<10$AU), the temperature can even become so high
(several thousand K) that photoevaporation becomes effective
\cite{Gorti04}.

The main effect of the UV radiation on the chemistry is through
photodissociation and photoionization of molecules and atoms in the
surface layers.  These processes remain effective until most photons
have been absorbed by the dust and the photorates have fallen below
those of other chemical reactions, typically at a depth of a few
magnitudes of extinction. The chemical structure of the outer disk has
been modeled by various groups in both the radial and vertical
direction (e.g., \cite{Willacy00,Aikawa02,Markwick02,vanZ03}).  In the
vertical direction, the chemical structure resembles that of a
photon-dominated region (PDR), with a transition from C$^+$ to C and
CO around $A_V\approx$1 mag and radicals such as CN and C$_2$H having
a significant abundance in the intermediate layers. In the cold
mid-plane ($<20$~K), most molecules are frozen out onto the
grains. This layered chemical structure is consistent with
observations of ions and radicals with abundance ratios that are
similar or higher than those found in PDRs (e.g.,
\cite{Dutrey97,Thi04}). If the UV radiation is energetic enough to
dissociate CO, isotope selective processes may occur leading to
fractionation of CO isotopes which can eventually be incorporated into
meteorites \cite{Lyons05}.  Finally, UV photons can affect the
chemistry through photodesorption of ices brought to the surface by
vertical mixing in the disk \cite{Willacy00}.

The above discussion illustrates the need for accurate
photodissociation cross sections to describe the disk chemistry.  In
this paper, photodissociation rates relevant for disks exposed to
different types of radiation fields are presented, including
calculations of cross sections for species not considered before.
Also, a critical evaluation of available cross sections at the Lyman
$\alpha$ wavelength is given. The resulting chemical structure is
illustrated for the transitional disk around the Herbig Ae/Be star HD
141569A.

\section{Photoprocesses in disks}
\label{photo}

The PDR structure of a disk differs from that of the more commonly
studied PDRs in molecular clouds \cite{Hollen97}.  First, the spectral
shape of the stellar radiation field can differ significantly from
that of the standard interstellar radiation field (ISRF), especially
for the cooler T Tauri stars. Figure 1 compares the standard ISRF cf.\
\cite{Draine78} with that of a 10000 K (typical Herbig Ae star) and a
4000 K (typical T Tauri star) radiation field, normalized to have the
same integrated intensity from 912--2050 \AA. Below 2000 \AA \ where
most molecules are photodissociated, the 4000 K field has orders of
magnitude lower intensity. The figure also includes a NEXTGEN
simulated spectrum of a B9.5 star \cite{Hauschildt99}, illustrating
that the intensity may be further reduced by stellar absorption lines
in the critical 912--1100 \AA \ where H$_2$ and CO are
photodissociated. These stellar lines become increasingly important at
lower stellar temperatures. Some T Tauri stars are known to have
excess UV emission above that of a 4000~K blackbody, either from a hot
boundary layer between the accretion disk and the star or from
chromospheric stellar activity, bringing the overall shape back closer
to that of the ISRF \cite{Calvet98,Kamp04b}. Specific resonance lines
like Lyman $\alpha$ can dominate the spectrum \cite{Bergin03}.  The
key chemical effect is that H$_2$ and CO are hardly photodissociated
by `cool' stars at 912--1100 \AA; only the general ISRF incident on
the disk sets up the H$\to$ H$_2$ and C$^+$$\to$C$\to$CO transitions
\cite{vanZ03}.  Other molecules with photodissociation channels
primarily at short wavelengths (e.g., N$_2$, CN) are equally affected.

Second, the UV field can be orders of magnitude higher than that
studied in molecular clouds. For example, at 15 AU from an A0 star,
the radiation is $10^7$ times that of the ISRF. For such high fluxes,
photorates can become comparable to dissociative recombination rates
so that photodissociation of ions needs to be included in the models.

Third, the grain sizes and properties may be very different from those
found in clouds. In particular, there is strong evidence from infrared
observations that grains in the upper disk layers have grown from
$\sim 0.1$ $\mu$m to a few $\mu$m in size regardless of the spectral
type of the star (e.g., \cite{Boekel05,Kessler06}).  Figure 2
illustrates this growth using recent Spitzer Space Telescope spectra
of the 10 and 20 $\mu$m silicate feature.  Larger grains have
significantly lower extinction at UV wavelengths. Also, PAHs may
have lower abundances and are detected toward only few T Tauri stars
\cite{Geers06}.

Fourth, the gas/dust mass ratio can differ from the interstellar
medium value of 100. Currently, it is unclear whether the (small) dust
particles disappear before the gas or vice versa. However, dust-free
gaps, rings and holes in disks have been found for some objects
including the HD 141569A disk (e.g., \cite{Augereau99,Clampin03}).
Some remaining gas can be present in these dust holes (e.g.,
\cite{Muzze03}).

Fifth, the density in the disk PDR varies by orders of magnitude, both
in vertical and radial direction, in contrast with the near constant
density assumed for clouds. If ices are released into the gas due to
vertical mixing followed by thermal or photodesorption, the initial
conditions for the PDR are different from those of standard
clouds. For example, the presence of enhanced gaseous H$_2$O has been
shown to modify the PDR chemistry for the case of protostellar
envelopes \cite{Stauber04}.

Finally, young stars are known to be powerful X-ray emitters (e.g.,
\cite{Feigel99}).  In addition to direct ionization, the UV radiation
resulting from the interaction of the secondary electrons with H$_2$
has a major effect on the chemistry \cite{Gredel89,Stauber05}.  This spectrum
consists of many discrete lines originating from the H$_2$ B and C
electronic states, with peaks at 1500--1600 \AA \ and shorter
wavelengths. Thus, many of the basic photodissociation processes
discussed here are also relevant for models which take X-rays into
account.

\section{Photodissociation cross sections and rates}
\label{photorates}

The photodissociation rate of a molecule can be computed from
$$ k_{pd}^{cont} = \int \sigma(\lambda) I(\lambda) d\lambda \ \ \ {\rm s^{-1}}
\eqno(1)$$
where $\sigma$ is the photodissociation cross section in cm$^2$ and
$I$ is the mean intensity of the radiation in photons cm$^{-2}$
s$^{-1}$ \AA$^{-1}$. Usually $\sigma$ is a broad, continuous function
with wavelength peaking close to the vertical excitation energy of the
electronic state involved in the process.  For photodissociation
initiated by line absorptions (e.g., predissociation), the rate
becomes
$$ k_{pd}^{line}= {{\pi e^2} \over m c^2} \lambda_{u\ell}^2 f_{u\ell} \eta_u
 I_{u\ell}
 \ \ \ {\rm s^{-1}} \eqno(2)$$
where $f_{u\ell}$ is the oscillator strength for absorption from lower
level $\ell$ to upper level $u$ and $\eta_u$ the dissociation
efficiency of state $u$, which lies between 0 and 1. The numerical
value of the factor $\pi e^2/mc^2$ is $8.85 \times 10^{-21}$ in the
adopted units with $\lambda$ in \AA. The total photodissociation rate
of a molecule is obtained by summing over all channels. 

Overviews of photodissociation cross sections and interstellar
photodissociation rates of astrophysically relevant molecules have
been given by \cite{Lee84,vanD88,Roberge91,Huebner92}.
Data on cross sections can be found in the chemical physics
literature, either from experiments (stable molecules) or theory
(radicals, ions).  These summaries include only limited data on the
photodissociation of small ions which are key to building carbon- and
oxygen species. Therefore, a literature search was performed and new
cross sections were calculated for several species.  Also, the first
critical evaluation of cross sections at the Lyman $\alpha$ wavelength
is presented. Finally, photodissociation rates are calculated for
different radiation fields and grain parameters.

\subsection{New photodissociation cross sections}

For small molecules and ions, quantum chemical calculations of the
potential energy curves and transition dipole moments combined with
dynamical calculations of the nuclear motions can provide accurate
photodissociation cross sections and oscillator strengths (see
\cite{Kirby88} for review).  All new calculations presented here were
performed with the MOLPRO set of programs \cite{Werner03}, using the
VTZ atomic orbital basis set \cite{Dunning71}. For neutral molecules,
diffuse $s$ and $p$ functions were added to allow for a proper
description of molecular Rydberg states.  Molecular orbitals were
generated in state-averaged complete active space (CAS)
calculations. In the ultimate contracted multi-reference
configuration-interaction (CI) calculation, typically 50 reference
configurations per symmetry were used. Thus, around 100000 contracted
configurations out of 300000 uncontracted configurations were
generated per symmetry.  Orbitals with an orbital energy below -15 eV
were kept doubly occupied in all CI calculations.

Typically, the lowest 5 electronic states of each symmetry were
calculated at the ground-state equilibrium geometry together with the
corresponding transition moments.  For diatomics, complete potential
energy curves were computed as well. Many of the excited states are
bound and the predisssociation efficiencies are often unknown.  In the
following, individual cases are discussed in more detail.

{\bf CH$_2^+$:} Accurate CI calculations by Theodorakopoulos \&
Petsalakis \cite{Theo91} show many dipole-allowed excited states below 13.6
eV. Their 1, 2 and 3$^2$B$_2$, 2, 3 and 4 $^2$A$_1$ and 2$^2$B$_1$
states have been included with $f$=0.008, 0.0001, 0.02, 0.01, 0.06,
0.05 and 0.03, respectively.

{\bf CH$_4^+$:} Detailed studies of the CH$_4^+$ photodissociation
processes starting from its lowest C$_{2v}$ state have been carried
out by van Dishoeck et al.\ \cite{vanD80} indicating that the higher excited
states are likely dissociative.  The oscillator strengths into the 2,
3 $^2$A$_1$ and 2$^2$B$_1$ states computed in this work are $f$=0.04,
0.04 and 0.08, respectively.

{\bf O$_2^+$:} Calculations show that all electronic states below 13.6
eV are bound \cite{Honjou78}. Our computed oscillator strengths to the
A$^2\Pi_u$ and 2$^2\Pi_u$ states ---the only dipole allowed states
below 13.6 eV and above the dissociation limit-- are only $f$=0.005
each.

{\bf CO$^+$:} CO$^+$ is strongly bound with a dissociation energy of
8.34 eV, but its higher excited D$^2\Pi$, G, E and F $^2\Sigma^+$
states are likely (pre-)dissociative \cite{Lavendy93}. The
calculated oscillator strengths to the D and G states are $f$=0.01 and
0.02, respectively. Similar values have been assumed for the higher
states.

{\bf HCO$^+$: }The excited electronic states and photodissociation
processes of HCO$^+$ have been studied in detail by Koch et al.\
\cite{Koch95a,Koch95b}.  HCO$^+$ is remarkably transparent at UV
wavelengths: the only dipole-allowed dissociative state is the
1$^1\Pi$ state around 11.5 eV  with a small cross section.

{\bf H$_2$O$^+$:} Calculations show that there are no dipole-allowed
dissociative electronic states below 13.6 eV so that the interstellar
H$_2$O$^+$ photodissociation rate is negligible.

{\bf SiO:} The electronic structure of SiO is similar to that of CO,
but with a lower dissociation energy of 8.26 eV implying that even the
lower Rydberg states can contribute to SiO photodissociation if they
are fully predissociated. Our computed oscillator strengths to the 3
$^1\Sigma^+$ and 2, 3, 4 and 5 $^1\Pi$ states are $f$=0.10, 0.32,
0.03, 0.11 and 0.10, respectively.

\subsection{Lyman $\alpha$ cross sections}

The Lyman $\alpha$ line can outshine the continuum radiation, so that
it is important to know whether a molecule can be photodissociated at
1216 \AA \ or not. Table 1 summarizes the cross sections for a number
of key molecules, including those that are known to have significant
abundances in the inner disk \cite{Markwick02}.  The table contains
references to the original experimental or theoretical work as well as
an assessment of the accuracy of the data, with the notation A
($\sigma$ known to better than 50\%), B ($\sigma$ uncertain up to a
factor of two) or C ($\sigma$ uncertain up to an order of magnitude).

The following molecules cannot be dissociated by Lyman $\alpha$
radiation: H$_2$, CO, N$_2$, CN, H$_3^+$, OH$^+$, H$_2$O$^+$ and
HCO$^+$.  An interesting case is formed by O$_2$. The absorption cross
section at 1215--1216 \AA \ is known to be small, of order
$(1-2)\times 10^{-20}$ cm$^2$ \cite{Ogawa75}, but there are resonances
into the $v'=0$ and 1 levels of the 2 $^3\Sigma_u^-$ state at 1245 and
1205 \AA, with $f=0.015$ each \cite{Buenker76}. If the Lyman $\alpha$
line is as wide as observed for some stars (1205--1230 \AA, $\pm$0.1 eV
\cite{Bergin03}), these states can make a small contribution.

Experimentally determined cross sections at Lyman $\alpha$ are
generally accurate to better than a factor of 2, especially when the
absorption is continuous.  For molecules for which only theoretical
calculations are available (e.g., CH, CH$_2$, C$_2$H, C$_3$, ...), the
cross sections at Lyman $\alpha$ are highly uncertain because the
energy level calculations are uncertain by $\sim$0.2 eV, which is not
accurate enough to determine whether there is an exact resonance of a
predissociative state at 1216 \AA.  For continuous absorption into a
directly repulsive state around 10.2 eV (e.g., OH), the computed
results are reliable, however.  For some molecules, no estimate is
possible because of lack of both experimental and theoretical data
around Lyman $\alpha$. This includes C$_3$H$_2$, SH, SH$^+$, CS and
SiO.

Only a few molecules can be photoionized by Lyman $\alpha$ (see Table
2 of \cite{vanD88} for ionization potentials). For species like NH$_3$
and NO, the photoionization cross sections or yields have been
measured so that the dissociation and ionization parts can be
separated.  In other cases such as CH$_3$, the branching ratio is
unknown.  SO has an ionization potential of 10.29 eV, just above the
Lyman $\alpha$ threshold, whereas CS$_2$ has 10.08 eV, just below.  Of
the major atoms, C, N, O, and S cannot be photoionized by Lyman
$\alpha$ whereas Mg, Si and Fe can.

\subsection{Photodissociation rates for Herbig Ae and T Tauri stars}

The database on photodissociation and ionization cross sections
assembled by van Dishoeck \cite{vanD88} has been used together with the above
new input to compute photodissociation rates for various radiation
fields. No attempt has been made to systematically review the chemical
physics literature since 1988 for molecules other than those discussed
in \S 3.1.  Minor updates are included for some species such as
CH$_2$, for which the cross sections of \cite{vanD96}
are adopted.  References to the sources for the cross sections are
given in Table 1 of \cite{vanD88}, which also includes the most
likely dissociation products. The resulting rates for the standard
ISRF cf.\ \cite{Draine78} are given in Tables 2 and 3.

In calculating the photodissociation rates, it has been assumed that
all states above the dissociation limit have 100\% dissociation
efficiency, thus providing a maximum value ($\eta_u=1$).  Even with
this assumption, the photodissociation of ions is on average
significantly slower than that of neutrals. Several ions (e.g.,
H$_3^+$, H$_2$O$^+$) do not even have any dipole-allowed dissociative
electronic transitions below 13.6 eV. Note that CH$^+$ dissociates
primarily into C + H$^+$ for the radiation fields considered here, not
C$^+$ + H as listed in UMIST99.  In contrast, OH$^+$ dissociates
primarily into O$^+$ + H.

The photodissociation rates for larger molecules are uncertain because
experimental data are sparse and incomplete. Moreover, often only
absorption cross sections are known, not the dissociation yields. For
large molecules, absorption is followed by internal conversion to
highly excited vibrational levels of the ground electronic state, only
some of which lead to dissociation.  Also, the products of the
photodissociation are largely unknown and can vary with wavelength;
see \cite{Millar06} for an illustrative example for the case of
CH$_3$OH. 

Tables 2 and 3 include the photodissociation and photoionization rates
for a $T_{BB}=10000$~K and a 4000~K blackbody radiation field. As
noted in \S 2, the shapes of the UV fields of Herbig Ae and T Tauri
stars are thought to lie in between these three extremes. Their
intensities have been normalized such that the integrated values from
912--2050 \AA \ are the same as those of the Draine \cite{Draine78} field,
$2.67\times 10^{-3}$ erg cm$^{-2}$ s$^{-1}$. The latter value is a
factor of 1.7 larger than that of the integrated Habing \cite{Habing68} field
of $1.6\times 10^{-3}$ erg cm$^{-2}$ s$^{-1}$ used in other
normalizations. The adopted dilution factors are $1.57\times 10^{-14}$
and $1.90\times 10^{-9}$ for 10000 and 4000 K, respectively. For
applications in disk chemistry, these rates need to be scaled to the
appropriate strength at a certain distance from the star.  The
photorates for (scaled) fields with $T_{BB}=20000-30000$~K are close
to those for the ISRF.

As shown previously for interstellar clouds exposed to cool stars
\cite{Spaans94}, the photorates can decrease by orders of magnitude if
a 4000 K blackbody is used. Species like H$_2$, CO, C, CN and N$_2$,
which are only dissociated or ionized at $<$1100 \AA, are most
affected (cf.\ Figure 1): their photorates become negligible for
$T_{BB}=4000$~K. In contrast, molecules with dissociation channels at
wavelengths as long as 3000 \AA \ (e.g., CH, HCO, O$_3$) can have
enhanced rates in the cooler fields with the adopted
normalization. Molecules which absorb over a broad wavelength range
(e.g., OH, H$_2$O, NO) vary by only a factor of a few between
$T_{BB}=30000$ and 4000 K.

Illustrative examples of the effect of Lyman $\alpha$ radiation on the
photorates in disks are given by \cite{Bergin03}. The
photodissociation and ionization efficiencies corresponding to Lyman
$\alpha$ absorption in the context of X-ray induced chemistry are
summarized by \cite{Lepp96} for selected species.

\subsection{Depth-dependence of photorates}

With depth into a cloud or disk, the UV radiation is attenuated due to
absorption and scattering by dust grains \cite{Roberge81}. For
typical 0.1 $\mu$m interstellar grains, the extinction, albedo and
scattering phase function are taken from \cite{Roberge91}. The
rates are then fitted to a single exponential decay with
$$ k_{pd} = k_{pd}^o \exp(-\gamma A_V) \ \ \ {\rm s^{-1}.} 
\eqno(3)$$
The fits of $\gamma$ are performed over the range $A_V=0-3$ mag and
are appropriate for a model in which the radiation is incident from one
side only. The absolute values of $\gamma$ vary by $\sim 10$\%
depending on the extinction range chosen for the fits; the relative values from
molecule to molecule are more accurate. Tables 2 and 3 contain the results
for the different radiation fields.

The values of $\gamma$ for the ISRF differ from those of van Dishoeck
\cite{vanD88} due to the adopted grain properties, which are in
between those of grain models 2 and 3 of \cite{Roberge81} used
previously.  Interestingly, the values of $\gamma$ do not change
dramatically with $T_{BB}$ for most species, even though there is a
general trend for $\gamma$ to become smaller with cooler fields (see
also \cite{Spaans94}).

When grains grow larger, the main effect on the grain parameters is
the smaller extinction at UV wavelengths, which becomes comparable to
that at visible wavelengths (e.g., \cite{Shen04}). The photorates have
been re-computed as a function of depth using the properties for
$\mu$m-sized ice-coated grains derived for the HD 141569A disk
\cite{Li03,Jonkheid06}.  For species such as C and CO which absorb
only at the shortest wavelengths, the exponents $\gamma$ are lowered
from $>3$ to $\sim$0.6 for the ISRF. For molecules which are
dissociated primarily at long wavelengths (e.g., HCO), $\gamma$ is
lowered from 1.1 to 0.47.  Other species have exponents $\gamma$ in
between these two values.  For $T_{BB}=4000$~K, the trends are
similar, with values of $\gamma$ between 0.6 and 0.38. Thus, the
differences between molecules in the depth dependence of the
photorates are minimized for large grains.

\section{A disk model example}
\label{disks}

In this work, the effects of different radiation fields and different
grain properties are taken into account explicitly by calculating at
each point in the disk the UV radiation field from the star as well as
the ISRF (either in 1+1D or in full 2D) and then computing the
photodissociation rates by integrating the cross sections multiplied
by the radiation intensity over wavelength \cite{vanZ03,Jonkheid06}.
Self-shielding of H$_2$ and CO as well as mutual shielding of CO and
its isotopes are included. X-rays are not included, nor is
time-dependence, vertical mixing or photodesorption of ices.

As an example, the disk around the Herbig Ae/Be star HD 141569A (B9.5
V, 22 L$_{\odot}$, 99 pc) is modelled. This $\sim$5 Myr old star has a
large $\sim$500 AU disk with a huge inner hole in the dust out to 150
AU and two dust rings at 185 and 325 AU seen in scattered light images
(e.g., \cite{Augereau99,Clampin03}\footnote{{see {\tt
http://hubblesite.org/newscenter/newsdesk/archive/releases/2003/02/}
for image}}).  The outer rings may be explained by tidal interactions
with two M-type companions, HD 141569B and C, but other explanations
such as a giant planet are not excluded. The origin of the inner hole
is unknown. The disk is optically thin to UV continuum radiation and
the grains have grown to at least $\mu$m size \cite{Li03}.
Nevertheless, gas is still present since the CO $J$=2--1 and 3--2
millimeter \cite{Dent05} and the $v=1-0$ infrared \cite{Brittain03}
lines have been detected. Moreover, PAHs have been seen, albeit at a
low level \cite{Sylvester96}. Thus, this disk provides a good
opportunity to investigate the effects of different UV radiation
fields and larger grains.

Detailed models to constrain the total gas mass in this disk have been
developed by Jonkheid et al.\ \cite{Jonkheid06}.  They use a 1+1D
approach, with 1D PDRs computed in the radial rather than vertical
direction for this optically thin disk. Both the radiation from the
star (see Figure~1) and the ISRF are included.
%The H$_2$ and CO columns are then determined in
%the vertical direction to compute the shielding of the ISRF. 
The size of the dust grains enters the chemistry in various ways: in
the UV extinction, in the H$_2$ formation rate, and through the
photoelectric heating efficiency \cite{Kamp01}.  The best-fitting
model to the CO millimeter lines has a total gas mass of 80 $M_{\rm
Earth}$, compared with a total dust mass of 2.2 $M_{\rm Earth}$ in
grains up to 1 cm. An important conclusion is that some gas must be
present in the inner hole to provide sufficient H$_2$ and CO
self-shielding in the radial direction.  The PAH abundance with
respect to total hydrogen is $\sim 10^{-10}$, but even at this low
abundance PAHs are an important site for H$_2$ formation.  Models
without PAHs require much larger gas masses \cite{Jonkheid06}.

Figures 3 and 4 present the radial and vertical distributions of
various molecules in the HD 141569A disk, both with and without the
ISRF. Because the stellar radiation field has few photons at
wavelengths below 1200 \AA, the photodissociation of CO and H$_2$ and
photoionization of C occurs mostly by the ISRF in the outer disk.  The
vertical slices show the typical PDR structure with atomic H and C$^+$
dominant on the surface, and the transition to H$_2$ taking place in
the intermediate layer. However, the CO column never becomes large
enough in the vertical direction for significant self-shielding so
that carbon stays in atomic form.  Other molecules like CH and C$_2$H
which are not self-shielding and can photodissociate over a large
wavelength range continue to be dissociated very rapidly by the
stellar radiation. Thus, even without the ISRF, the CO abundance stays
low since its precursor molecules have very low abundances. Neutral
atomic carbon is the dominant form of carbon since there are few
ionizing photons. The predicted peak line intensity of $\sim$1 K
should be readily detectable by submillimeter telescopes
\cite{Jonkheid06}. Failure to detect this line may imply that the
observed CO is not a relic of the interstellar cloud from which HD
141569A formed.
%a non-detection would likely imply the presence of
%carbon ionizing radiation not accounted for in this model.

\section{Conclusions}

The main results of this work are as follows.

\begin{itemize}

\item{New photodissociation cross sections are presented for several species
not considered before. Also, the first critical evaluation of cross sections
at Lyman $\alpha$ is given.}

\item{The photodissociation of ions is potentially important in (inner)
disk chemistry where the radiation field can be $10^7$ times more
intense than the standard ISRF. However, the photorates of many ions,
especially those containing oxygen, are found to be slow compared with
those of neutrals.}

\item{Photorates have been computed for a range of radiation fields
appropriate for Herbig Ae/Be and T Tauri stars. Molecules such as
H$_2$, CO, N$_2$ and CN, which are photodissociated only below 1200
\AA, have rates that are more than 5 orders of magnitude decreased for
a $T_{BB}=4000$ K blackbody field.}

\item{The exponent $\gamma$ characterizing the depth-dependence of the
photorates decreases with $T_{BB}$ but the effect is not large in the
4000--30000 K range for most species. For larger $\mu$m-sized grains,
the depth dependence of the photorates becomes much shallower, with
$\gamma$ falling in a narrow range of 0.4--0.6 for all species.}

\item{The example of the HD 141569A disk illustrates that both the shape
of the stellar radiation field and the size of the grains affect the
chemistry.  Both need to be treated correctly to derive quantitative
conclusions about the gas mass and chemistry from observed lines.
Even a small amount of PAHs or small grains can significantly affect
dissociation and ionization rates, as well as the H$_2$ formation
rate.}

\end{itemize}

\section{Acknowledgments}

We are grateful to the Spitzer `Cores to Disks' team, in particular
J. Kessler-Silacci and V. Geers, for providing observational data.
The HD 141569A modeling is performed in collaboration with J.C.\
Augereau and I. Kamp. Astrochemistry in Leiden is supported by a
Spinoza grant from the Netherlands Organization for Scientific
Research (NWO).

\clearpage
\begin{table}
\begin{center}
\caption{\label{tab1}Photodissociation and ionization
cross sections at Lyman $\alpha$
1216 \AA$^a$}
\begin{tabular}{llcl}
\hline
Species & \ \ $\sigma_{\rm pd}$ & Accuracy$^b$ & Ref\\
        & \ \ (cm$^{-2}$) & \\
\hline
CH                   & 5.0(-20)         & C & \cite{vanD87} \ T \\
CH$_2$               & 5.0(-20)         & C & \cite{vanD96} \ T \\
CH$_4$               & 1.8(-17)        & A & \cite{Lee83} \ E \\
C$_2$                & 5.0(-18)        & B & \cite{Pouilly83} \ T \\
C$_3$                & 1.0(-18)        & C & \cite{Romelt78} \ T \\
C$_2$H               & 1.0(-18)        & C & \cite{Shih79} \ T \\
C$_2$H$_2$           & $\geq$4(-17)$^c$   & B & \cite{Suto84} \ E \\
C$_4$H$_2$           & 3.5(-17)        & B & \cite{Okabe81} \ E \\
OH                   & 1.8(-18)        & B & \cite{vanD84} \ T \\
H$_2$O               & 1.2(-17)        & A & \cite{Lee86} \ E \\
O$_2$                & 1.0(-20)$^b$     & C & \cite{Ogawa75} \ E \\
CO$_2$               & 6.1(-20)         & A & \cite{Yoshino96} \ E \\
H$_2$CO              & 1.0(-17)         & B & \cite{Suto86} \ E \\
CH$_3$OH             & 1.4(-17)         & A & \cite{Nee85} \ E \\
NH                   & 1.0(-18)         & B & \cite{Kirby91} \ T \\
NH$_3$               & 1.0(-17)        & A & \cite{Suto83} \ E \\
HCN                  & 3.0(-17)        & A & \cite{Lee80} \ E \\
HC$_3$N              & 2.5(-17)        & B & \cite{Connors74} \ \ E \\
CH$_3$CN             & 2.0(-17)        & A & \cite{Suto85} \ E \\
NO                   & 4.0(-19)        & B & \cite{Guest81} \ E \\ 
H$_2$S               & 3.3(-17)        & B & \cite{Lee84} \ E \\
SO                   & 1.0(-16)        & C & \cite{Nee86} \ E \\
SO$_2$               & 3.0(-17)        & B & \cite{Lee84} \ E \\
OCS                  & 1.5(-17)        & A & \cite{Lee84} \ E \\
CS$_2$               & 2.5(-17)        & B & \cite{Lee84} \ E \\ [3mm]
Mg p.i.              & 3.0(-19)        & A & \cite{vanD88} \\
Si p.i.              & 3.0(-17)        & A & \cite{vanD88}   \\
Fe p.i.              & 6.2(-19)        & A & \cite{vanD88} \\
NH$_3$ p.i.          & 2.0(-18)        & B & \cite{Watanabe65} \ E \\         
NO p.i.              & 1.6(-18)        & B & \cite{Guest81} \ E \\
CS$_2$ p.i.          & 2.0(-16)        & B & \cite{Lee84} \ E \\
\hline
\end{tabular}
\end{center}
$^a$ See text for list of species that cannot be dissociated or ionized by Lyman $\alpha$ \\
$^b$ See text for discussion \\
$^c$ $f$=0.013 \\
\end{table}

\clearpage
\begin{table}
\begin{center}
\caption{\label{tab2}Photodissociation rates for 
various radiation fields$^{a,b}$}
\begin{tabular}{lllllll}
\hline
Species & \multicolumn{3}{c}{$k_{pd}^o$ (s$^{-1}$)} & 
\multicolumn{3}{c}{$\gamma$}\\ \cline{2-7} 
&ISRF$^c$ & 10000 K$^d$ & 4000 K$^d$ & ISRF & 10000 K & 4000 K \\ \hline
H$_2^+$      & 5.7(-10) & 1.9(-10) & 2.9(-11) & 2.37 & 2.14 & 1.99 \\
CH           & 9.2(-10) & 2.0(-9)  & 1.2(-7)  & 1.72 & 1.49 & 1.28 \\
CH$^+$       & 3.3(-10) & 3.5(-11) & 4.8(-10) & 2.94 & 1.78 & 1.31 \\
CH$_2$       & 5.8(-10) & 1.2(-9)  & 2.1(-9)  & 2.02 & 2.02 & 2.12 \\
CH$_2^+$     & 1.4(-10) & 7.4(-11) & 2.6(-11) & 2.21 & 1.91 & 1.88 \\
CH$_3$       & 2.7(-10) & 2.5(-10) & 8.2(-10) & 2.27 & 2.24 & 2.32 \\
CH$_4$       & 1.2(-9)  & 2.2(-10) & 1.2(-12) & 2.59 & 2.45 & 2.29 \\
CH$_4^+$     & 2.8(-10) & 4.2(-11) & 1.3(-13) & 2.71 & 2.58 & 2.48 \\
C$_2$        & 2.4(-10) & 4.1(-11) & 3.2(-13) & 2.57 & 2.36 & 2.25 \\
C$_2$H       & 5.2(-10) & 1.9(-10) & 7.2(-12) & 2.30 & 2.16 & 2.10 \\
C$_2$H$_2$   & 3.3(-9)  & 1.2(-9)  & 1.3(-10) & 2.27 & 2.12 & 1.97 \\
C$_2$H$_4$   & 3.0(-9)  & 2.2(-9)  & 5.2(-10) & 2.10 & 1.96 & 1.90 \\
C$_3$        & 3.8(-9)  & 2.9(-9)  & 2.0(-10) & 2.08 & 2.07 & 2.06 \\
$c-$C$_3$H$_2$ & 1.9(-9) & 1.7(-9) & 9.2(-10) & 2.07 & 2.06 & 2.10 \\
OH           & 3.9(-10) & 1.8(-10) & 1.3(-10) & 2.24 & 2.00 & 1.67 \\
OH$^+$       & 1.1(-11) & 7.8(-13) & 5.8(-13) & 3.50 & 2.80 & 1.75 \\
H$_2$O       & 8.0(-10) & 4.3(-10) & 1.2(-10) & 2.20 & 1.97 & 1.90 \\
O$_2$        & 7.9(-10) & 4.9(-10) & 4.5(-11) & 2.13 & 2.05 & 1.97 \\
O$_2^+$      & 3.5(-11) & 3.6(-11) & 1.0(-11) & 2.02 & 1.92 & 1.90 \\
HO$_2$       & 6.7(-10) & 1.9(-9)  & 1.2(-8)  & 2.12 & 2.08 & 1.99 \\
H$_2$O$_2$   & 9.5(-10) & 4.3(-10) & 1.4(-10) & 2.28 & 2.07 & 1.97 \\
O$_3$        & 1.9(-9)  & 5.4(-9)  & 1.5(-7)  & 1.85 & 1.69 & 1.57 \\
CO           & 2.0(-10) & 1.5(-11) & 1.4(-15) & 3.53 & 3.47 & 3.24 \\
CO$^+$       & 1.0(-10) & 2.2(-11) & 1.2(-13) & 2.52 & 2.43 & 2.32 \\
CO$_2$       & 8.9(-10) & 9.0(-11) & 1.2(-12) & 3.00 & 2.53 & 2.00 \\
HCO          & 1.1(-9)  & 2.5(-9)  & 3.5(-6)  & 1.09 & 1.09 & 0.81 \\
HCO$^+$      & 5.4(-12) & 4.5(-13) & 7.9(-17) & 3.32 & 3.32 & 3.32 \\
H$_2$CO      & 1.0(-9)  & 6.7(-10) & 1.8(-10) & 2.16 & 1.99 & 1.90 \\
CH$_3$OH     & 1.4(-9)  & 5.9(-10) & 7.0(-11) & 2.28 & 2.07 & 1.95 \\
\hline
\end{tabular}
\end{center}
\end{table}

\clearpage
\begin{table}
\begin{center}
\caption{\label{tab2}Photodissociation rates for various radiation fields$^{a,b}$
(Table 2 cont'd)}
\begin{tabular}{lllllll}
\hline
Species & \multicolumn{3}{c}{$k_{pd}^o$ (s$^{-1}$)} & 
\multicolumn{3}{c}{$\gamma$}\\ \cline{2-7} 
&ISRF$^c$ & 10000 K$^d$ & 4000 K$^d$ & ISRF & 10000 K & 4000 K \\ \hline
NH           & 5.0(-10) & 1.6(-10) & 3.0(-12) & 2.33 & 2.24 & 2.12 \\
NH$^+$       & 5.4(-11) & 1.8(-10) & 8.4(-9)  & 1.64 & 1.52 & 1.51 \\
NH$_2$       & 7.5(-10) & 1.0(-9)  & 5.7(-10) & 2.00 & 1.90 & 1.89 \\
NH$_3$       & 1.2(-9)  & 1.0(-9)  & 1.1(-9)  & 2.12 & 1.99 & 2.00 \\
N$_2$        & 2.3(-10) & 1.4(-11) & 3.0(-16) & 3.88 & 3.89 & 3.87 \\
NO           & 4.7(-10) & 4.3(-10) & 2.9(-10) & 2.12 & 1.96 & 1.94 \\
NO$_2$       & 1.4(-9)  & 1.0(-9)  & 3.4(-10) & 2.12 & 1.97 & 1.92 \\
N$_2$O       & 1.9(-9)  & 4.8(-10) & 2.0(-11) & 2.44 & 2.32 & 2.02 \\
CN           & 2.9(-10) & 2.1(-11) & 2.0(-15) & 3.54 & 3.49 & 3.23 \\
HCN          & 1.6(-9)  & 2.5(-10) & 3.7(-12) & 2.69 & 2.44 & 2.02 \\
HC$_3$N      & 5.6(-9)  & 3.0(-9)  & 2.5(-10) & 2.16 & 2.12 & 2.12 \\
CH$_3$CN     & 2.5(-9)  & 4.8(-10) & 8.5(-12) & 2.58 & 2.38 & 2.01 \\
SH           & 9.8(-10) & 1.3(-9)  & 1.6(-8)  & 2.04 & 1.85 & 1.34 \\
SH$^+$       & 2.5(-10) & 3.4(-10) & 4.0(-8)  & 1.66 & 1.30 & 1.29 \\
H$_2$S       & 3.1(-9)  & 2.0(-9)  & 3.2(-9)  & 2.27 & 2.12 & 2.16 \\
CS           & 9.8(-10) & 2.7(-10) & 3.7(-12) & 2.43 & 2.33 & 2.14 \\
CS$_2$       & 6.1(-9)  & 1.3(-8)  & 2.0(-8)  & 2.06 & 2.02 & 2.03 \\
OCS          & 3.7(-9)  & 3.1(-9)  & 7.0(-10) & 2.07 & 1.98 & 1.94 \\
SO           & 4.2(-9)  & 4.4(-9)  & 9.4(-9)  & 2.37 & 2.18 & 2.16 \\
SO$_2$       & 1.9(-9)  & 7.4(-10) & 2.8(-10) & 2.38 & 2.11 & 1.94 \\
SiH          & 2.8(-9)  & 1.4(-8)  & 8.0(-7)  & 1.59 & 1.55 & 1.22 \\
SiH$^+$      & 2.7(-9)  & 1.0(-8)  & 3.3(-6)  & 1.21 & 1.13 & 1.11 \\
SiO          & 1.6(-9)  & 5.6(-10) & 9.9(-12) & 2.28 & 2.21 & 2.19 \\
\hline
\end{tabular}
\end{center}
$^a$ See van Dishoeck \cite{vanD88} for products and 
references to cross section 
data \\
$^b$ H$_3^+$, HeH$^+$ and H$_2$O$^+$ cannot be photodissociated by
radiation with $\lambda > 912$ \AA, so $k_{pd}^o$=0 \\
$^c$ ISRF according to \cite{Draine78} with extension at $>$2000 \AA \ of
\cite{vanD82} \\
$^d$ Scaled blackbody radiation field with temperature $T_{BB}$ (see text)
\end{table}

\clearpage
\begin{table}
\begin{center}
\caption{\label{tab3}Photoionization rates for various radiation fields$^a$
(Table 3)}
\begin{tabular}{lllllll}
\hline
Species & \multicolumn{3}{c}{$k_{pd}^o$ (s$^{-1}$)} &
\multicolumn{3}{c}{$\gamma$}\\ \cline{2-7}
&ISRF$^b$ & 10000 K$^c$ & 4000 K$^c$ & ISRF & 10000 K & 4000 K \\ \hline
C           & 3.1(-10) & 2.5(-11) & 4.2(-15) & 3.33 & 3.27 & 3.10 \\
Mg          & 7.9(-11) & 5.9(-11) & 6.9(-12) & 2.08 & 2.00 & 1.96 \\
Si          & 3.1(-9)  & 1.2(-9)  & 4.1(-11) & 2.27 & 2.17 & 2.09 \\
S           & 6.0(-10) & 5.9(-11) & 4.0(-14) & 3.08 & 2.95 & 2.76 \\
Fe          & 2.8(-10) & 1.3(-10) & 5.8(-12) & 2.20 & 2.14 & 2.05 \\ [3mm]
CH          & 7.6(-10) & 6.4(-11) & 1.6(-14) & 3.28 & 3.20 & 2.97 \\
CH$_4$      & 6.8(-12) & 4.7(-13) & 5.0(-18) & 3.94 & 3.94 & 3.93 \\
C$_2$       & 4.1(-10) & 2.6(-11) & 1.0(-15) & 3.81 & 3.81 & 3.78 \\
C$_2$H$_2$  & 3.3(-10) & 2.4(-11) & 2.4(-15) & 3.52 & 3.49 & 3.33 \\
C$_2$H$_4$  & 4.1(-10) & 3.5(-11) & 1.1(-14) & 3.21 & 3.11 & 2.91 \\
C$_2$H$_6$  & 2.3(-10) & 1.5(-11) & 7.6(-16) & 3.74 & 3.73 & 3.55 \\
O$_2$       & 7.6(-11) & 4.8(-12) & 1.2(-16) & 3.87 & 3.87 & 3.85 \\
H$_2$O      & 3.1(-11) & 2.0(-12) & 3.6(-17) & 3.90 & 3.90 & 3.88 \\
NH$_3$      & 2.8(-10) & 2.6(-11) & 1.1(-14) & 3.12 & 3.04 & 2.86 \\
NO          & 2.6(-10) & 2.9(-11) & 5.4(-14) & 2.93 & 2.71 & 2.42 \\
NO$_2$      & 1.5(-10) & 1.2(-11) & 2.9(-15) & 3.33 & 3.22 & 2.87 \\
N$_2$O      & 1.7(-10) & 1.1(-11) & 1.4(-16) & 3.93 & 3.93 & 3.92 \\
H$_2$S      & 7.3(-10) & 7.2(-11) & 3.5(-14) & 3.09 & 3.01 & 2.86 \\
CS$_2$      & 1.7(-9)  & 1.5(-10) & 8.3(-14) & 3.16 & 3.02 & 2.77 \\
OCS         & 6.9(-10) & 1.4(-11) & 8.5(-15) & 3.35 & 3.29 & 3.10 \\
H$_2$CO     & 4.8(-10) & 4.1(-11)  & 1.2(-14) & 3.21 & 3.13 & 2.96 \\
\hline
\end{tabular}
\end{center}
$^a$ See van Dishoeck \cite{vanD88} for references to cross section 
data \\
$^b$ See footnote $c$ Table 2 \\
$^c$ See footnote $d$ Table 2 
\end{table}

%Please compile a list of all figure captions on a separate page:
\clearpage
\begin{list}{}{\leftmargin 2cm \labelwidth 1.5cm \labelsep 0.5cm}

\item[{\bf Fig. 1}] Comparison of the interstellar radiation field
(ISRF) according to \cite{Draine78} (with the extension by
\cite{vanD82} for $>$2000 \AA) with a 10000 K (dashed) and 4000 K
(dotted) blackbody scaled to have the same integrated intensity from
912--2050 \AA. The scaled NEXTGEN model radiation field a B9.5 star
\cite{Hauschildt99} (as appropriate for HD 141569A) is included as
well (dash-dotted).

\item[{\bf Fig. 2}] Evidence for grain growth in disks around T Tauri
stars.  Top: Spitzer Space Telescope observations of the 10 and 20
$\mu$m silicate Si-O stretching and O-Si-O bending mode features of
two T Tauri stars, shifted by +0.4 and +0.2 for clarity.  Bottom:
normalized absorption efficiencies $Q_{\rm abs}$ for models of
spherical amorphous olivines with various sizes, calculated using the
distribution of hollow spheres (DHS) method of \cite{Min05}.  The
models are offset by +0.4, +0.25 and +0.1, respectively.  Figure based
on \cite{Kessler06}.

\item[{\bf Fig. 3}] Left: Radial distribution in the midplane of the
HD 141569A disk of the temperature and density (top), and various
chemical species (middle and bottom). Right: Vertical slice at $R=300$
AU. This model includes both the stellar radiation (Figure 1) and the ISRF.

\item[{\bf Fig. 4}] As Figure 3, but without the ISRF

\end{list}

\clearpage

\begin{figure}[ht]
  \begin{center}
   \includegraphics[width=10cm,angle=-90]{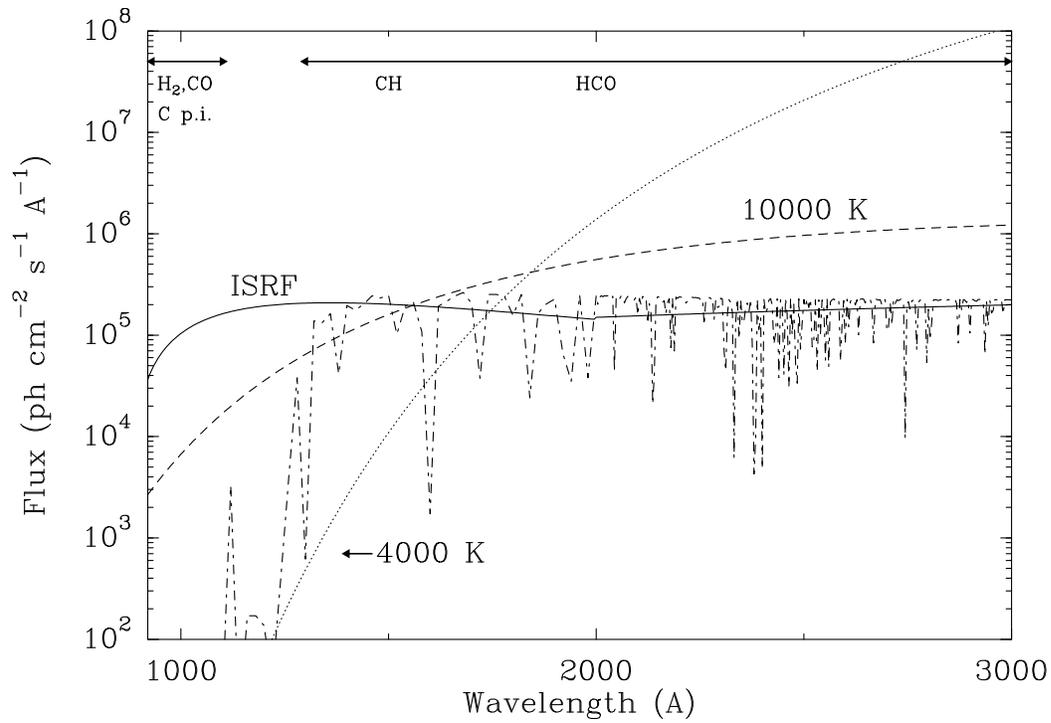}
  \caption{
Comparison of the interstellar radiation field
(ISRF) according to \cite{Draine78} (with the extension by
\cite{vanD82} for $>$2000 \AA) with a 10000 K (dashed) and 4000 K
(dotted) blackbody scaled to have the same integrated intensity from
912--2050 \AA. The scaled NEXTGEN model radiation field a B9.5 star
\cite{Hauschildt99} (as appropriate for HD 141569A) is included as
well (dash-dotted).}
  \end{center}
\end{figure}

\clearpage

\begin{figure}[ht]
  \begin{center}
   \includegraphics[width=15cm,angle=-90]{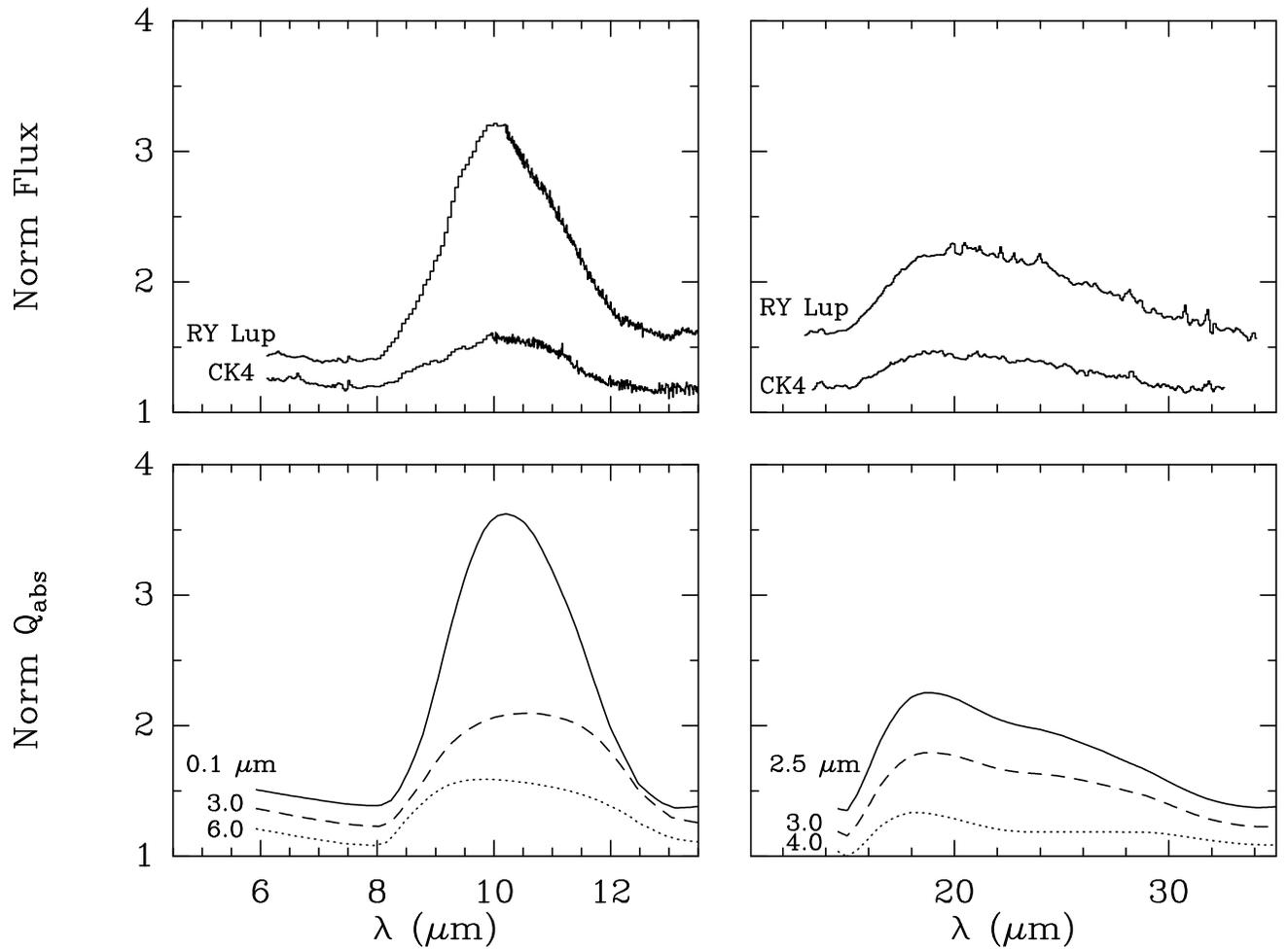}
   \caption{
Evidence for grain growth in disks around T Tauri
stars.  Top: Spitzer Space Telescope observations of the 10 and 20
$\mu$m silicate Si-O stretching and O-Si-O bending mode features of
two T Tauri stars, shifted by +0.4 and +0.2 for clarity.  Bottom:
normalized absorption efficiencies $Q_{\rm abs}$ for models of
spherical amorphous olivines with various sizes, calculated using the
distribution of hollow spheres (DHS) method of \cite{Min05}.  The
models are offset by +0.4, +0.25 and +0.1, respectively.  Figure based
on \cite{Kessler06}. }
  \end{center}
\end{figure}

\clearpage

\begin{figure}[ht]
  \begin{center}
   \includegraphics[width=15cm]{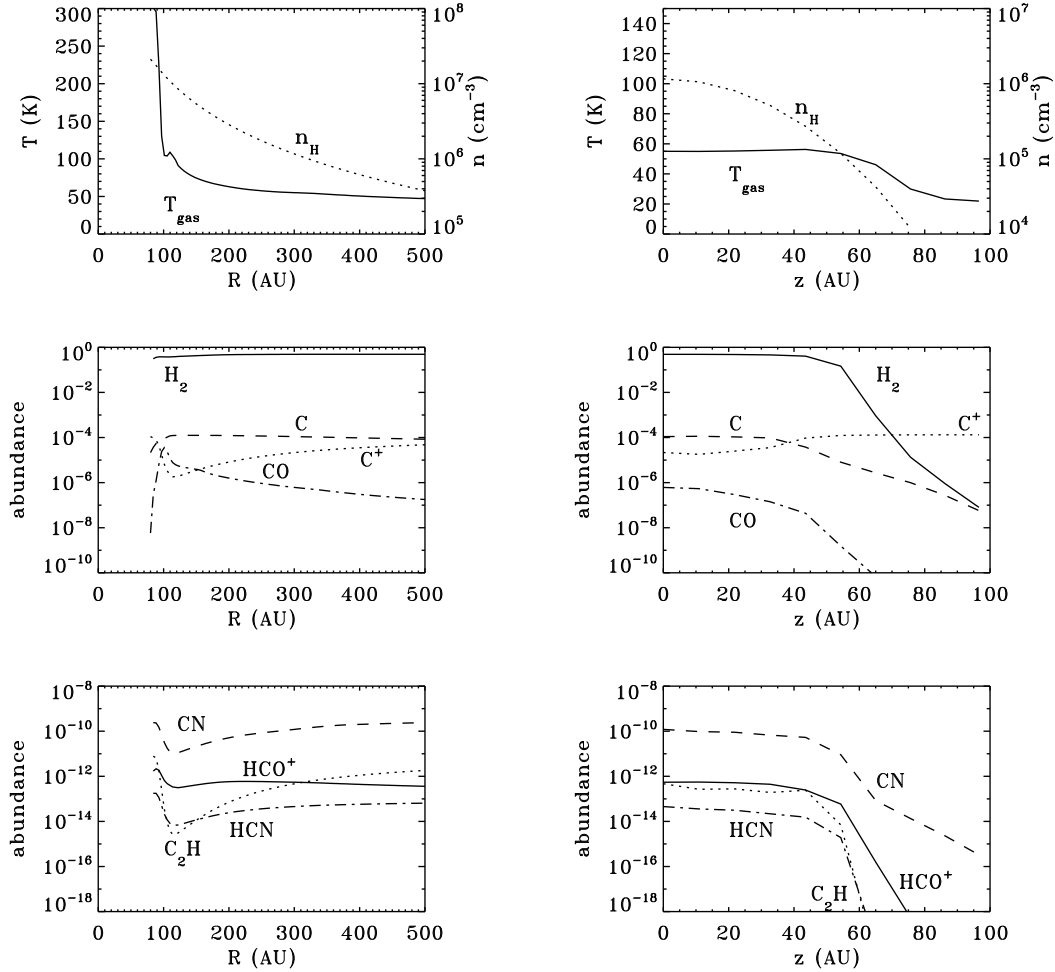}
   \caption{
Left: Radial distribution in the midplane of the
HD 141569A disk of the temperature and density (top), and various
chemical species (middle and bottom). Right: Vertical slice at $R=300$
AU. This model includes both the stellar radiation (Figure 1) and the ISRF.
}
  \end{center}
\end{figure}

\clearpage

\begin{figure}[ht]
  \begin{center}
   \includegraphics[width=15cm]{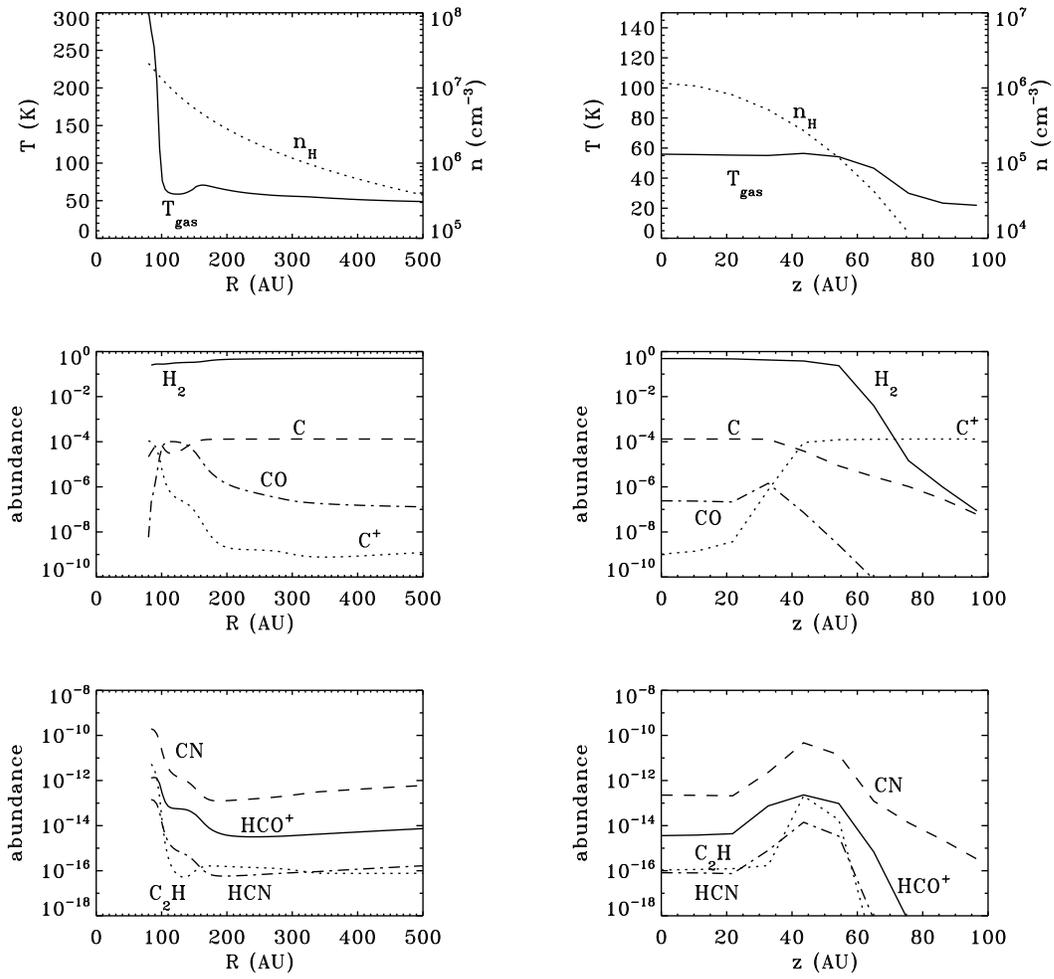}
   \caption{As Figure 3, but without the ISRF.}
  \end{center}
\end{figure}

\end{document}